\documentclass[%
 reprint,
 amsmath,amssymb,
 aps,
]{revtex4-2}

\usepackage{graphicx}
\usepackage{dcolumn}
\usepackage{bm}

\newtheorem{proposition}{Proposition}
\renewcommand{\t}{\mathsf{ t}}
\newcommand{\sh}{\mathsf{ h}} 
\renewcommand{\i}{i}

\newcommand{\eps}{\epsilon}
\newcommand{\g}{\mathfrak{g}}

\newcommand{\what}{\widehat}

\newcommand{\til}{\widetilde}

\newcommand{\CP}{\mathbb{CP}}

\newcommand{\norm}[1]{\left\| #1 \right\|}

\newcommand{\op}{\operatorname}

\newcommand{\mbb}{\mathbb}

\newcommand{\mc}{\mathcal}

\newcommand{\ip}[1]{\left\langle #1 \right\rangle}
\newcommand{\abs}[1]{\left| #1 \right|}

\newcommand{\R}{\mbb R}
\renewcommand{\d}{\mathrm{d}}

\newcommand{\half}{\tfrac{1}{2}}

\begin{document}

\preprint{APS/123-QED}

\title{On the associativity of one-loop corrections to the celestial OPE}

\author{Kevin Costello}
 \email{kcostello@perimeterinstitute.ca}
\affiliation{Perimeter Institute for Theoretical Physics, Waterloo, ON, CA
}%


\author{Natalie M. Paquette}
\email{npaquett@uw.edu}
\affiliation{Department of Physics, University of Washington, Seattle, WA
}%
%

\date{\today}

\begin{abstract}
	There has been recent interest in the question of whether QCD collinear singularities can be viewed as the OPE of a two-dimensional CFT. We analyze a version of this question for the self-dual limit of pure gauge theory (incorporating states of both helicities).   We show that the known one-loop collinear singulaties do not form an associative chiral algebra.  The failure of associativity can be traced to a novel gauge anomaly on twistor space.   We find that associativity can be restored for certain gauge groups if we introduce an unusual axion, which cancels the twistor space anomaly by a Green-Schwarz mechanism. Alternatively, associativity can be restored for some gauge groups with carefully chosen matter.  
\end{abstract}

\maketitle


\section{\label{sec:intro}Introduction}
The celestial holography program (see the recent reviews \cite{Strominger:2017zoo, Raclariu:2021zjz, Pasterski:2021rjz} and references therein) suggests, among other things, that collinear singularities in the scattering amplitudes of gauge theory and gravity are controlled by a CFT.   This is known to be true at tree level \cite{Guevara:2021abz}, and in the beautiful paper \cite{Ball:2021tmb} it was shown to persist to one-loop level in one formulation of self-dual gravity and gauge theory.

In this Letter we analyze a different formulation of this question for self-dual gauge theory, and find a different result.   In our work, we define self-dual gauge theory to include states of both helicities, but with the Lagrangian $\int B F(A)_-$.   This is in contrast to \cite{Ball:2021tmb}, where only states of positive helicity are considered.   With our definition, which is common in twistor studies \cite{Mason:2005zm},  self-dual gauge theory can be deformed to QCD by adding the operator $\half \op{tr}(B^2)$.  

We analyze collinear singularities that appear not just in amplitudes, but in \emph{form factors}.   Form factors are scattering amplitudes in the presence of a local operator.

We say a collinear singularity is \emph{universal} if it appears in the same way in all form factors.   Universal collinear singularities in self-dual gauge theory capture certain collinear singularities in QCD.  This is because certain form factors of self-dual gauge theory for the operator $\op{tr}(B^2)$ are the same as certain QCD amplitudes (e.g.\ at one loop they compute QCD amplitudes with one negative helicity gluon). 

We ask the question: \emph{do universal collinear singularities in self-dual gauge theory form a CFT}? We find that the answer is \emph{no} : associativity of the OPE fails at one loop \footnote{We emphasize that there is no contradiction with the work of \cite{Ball:2021tmb}, because of the different ways in which we define self-dual gauge theory, and because of our use of form factors in addition to amplitudes. }. 

We can trace the failure of associativity to an anomaly on twistor space. In \cite{Costello:2022wso} we studied universal collinear singularities of self-dual gauge theory, using a twistor space analysis.   The twistor uplift of self-dual gauge theory is holomorphic BF theory \cite{Ward:1977ta}. It was shown in \cite{Costello:2021bah} that holomorphic BF theory has a one-loop gauge anomaly on twistor space, that can be cancelled by the introduction of an additional field.  The cancellation takes the form of a Green-Schwarz mechanism, and holds if the gauge group is $SU(2)$, $SU(3)$, $SO(8)$ or an exceptional group. 

On space-time this additional field becomes an axion-like field (which we simply refer to as an axion hereafter), with a fourth-order kinetic term \footnote{It is interesting to note that a scalar dilaton field with a fourth-order kinetic term has been used to cancel conformal anomalies in four-dimensions \cite{FT, FT2, Riegert, Komargodski:2011vj}. We thank Z. Komargodski for bringing this to our attention. We will continue to call our field the axion because it couples to a topological term in the gauge field. }:
\begin{multline} 
	\int \op{tr}(B F(A)_-) - \half \int (\square \rho)^2   \\ - \frac{\sqrt{5}  2 \sh^\vee } { \sqrt{2 (\dim \g + 2)}     8 \pi \sqrt{3}  }  \int \rho \op{tr}(F(A) \wedge F(A)). \label{eqn:axion}
\end{multline}
 where $\sh^\vee$ is the dual Coxeter number.

In \cite{Costello:2022wso} we showed on abstract grounds that, when we cancel the twistor space anomaly, the universal collinear singularities have the structure of a chiral algebra.  

We find (by an explicit computation) that associativity of QCD collinear singularities is \emph{restored} when we add the axion.   This is a Green-Schwarz mechanism for associativity of the collinear singularities.

Conversely, once we introduce the axion, associativity of the OPE forces the collinear singularities in the gauge sector to have certain one-loop corrections. These include the standard one-loop QCD correction.  In this way, we find a purely chiral-algebraic computation of the standard one-loop collinear singularities.  

Finally, we leverage associativity to find a remarkably simple formula for the one-loop amplitudes of QCD with an axion (or of QCD without an axion, but with carefully chosen matter and gauge group).

\section{The chiral algebra}
We will start by reviewing the tree-level chiral algebra encoding collinear singularities in self-dual gauge theory.  Our analysis uses analytically-continued momenta, and so works in any signature.  As is standard, in the spinor-helicity formalism states are expressed in terms of spinors $\lambda_{\alpha}$, $\til{\lambda}^{\dot{\alpha}}$.    The chiral algebra lives on a copy of $\CP^1$ with homogeneous coordinates $(\lambda_1:\lambda_2)$.  We will use a coordinate $z$ corresponding to $(\lambda_1:\lambda_2)  =(1:z)$.

The chiral algebra is generated by two towers of states $J_a[r,s](z)$, $\til{J}_a[r,s](z)$,  corresponding to particles of positive and negative helicity, respectively.  We can arrange these into generating functions 
\begin{equation} 
	\begin{split} 
		J_a[\til{\lambda}](z) &=	\sum \omega^{r+s} \frac{1}{r! s!} (\til{\lambda}^{\dot{1}})^r (\til{\lambda}^{\dot{2}})^s  J_a[r,s](z) \\
		\til{J}[\til{\lambda}](z) &=  \sum \omega^{r+s} \frac{1}{r! s!} (\til{\lambda}^{\dot{1}})^r (\til{\lambda}^{\dot{2}})^s  \til{J}_a[r,s](z).
	\end{split}
\end{equation}
These generating functions correspond to gauge theory states of positive and negative helicity, with momenta encoded in the spinors $\til{\lambda}$ and $\lambda = (1,z)$.   Because we are expanding in powers of the energy $\omega$, these chiral algebra states should be thought of as soft modes. Precisely, we have $\mathcal{O}_a^-(z) = {\omega}\til{J}_a[\til{\lambda}](z), \mathcal{O}_a^+(z) = {1 \over \omega}J_a[\til{\lambda}](z)$, where $\mathcal{O}^{\pm}$ are the positive and negative helicity hard gluon operators dual to momentum eigenstates of energy $\omega$. 

In expressing the OPE, we write
\begin{equation} 
	\begin{split} 
		\ip{ij} &= 2 \pi \i (z_i - z_j) \\
		[ij] &=- 2 \eps_{\dot{\alpha}\dot{\beta}} \til{\lambda}_i^{\dot \alpha} \til{\lambda}_j^{\dot \beta}.  
	\end{split}
\end{equation}
In the supplemental material we provide more details on these conventions. The normalization of $[ij]$ is in order to match standard conventions where $\ip{ij}[ij] = 2 p_i \cdot p_j$. 

The tree-level OPE was derived in \cite{Costello:2022wso} using twistor space methods, but also matches the standard \cite{Mangano:1987kp} tree-level splitting amplitudes. The tree-level OPE is
\begin{equation} 
	\begin{split} 
		J_a[\til{\lambda}_1] (z_1) J_b[\til{\lambda}_2](z_2) &\sim f_{ab}^c \frac{1}{\ip{12}} J_c[\til{\lambda}_1 + \til{\lambda}_2]  (z_1) 	\\
		J_a[\til{\lambda}_1] (z_1) \til{J}_b[\til{\lambda}_2](z_2) &\sim f_{ab}^c \frac{1}{\ip{12}} \til{J}_c[\til{\lambda}_1 + \til{\lambda}_2]  (z_1).	
	\end{split}
\end{equation}
These two OPEs correspond to splitting amplitudes $+ \mapsto ++$ and $- \mapsto +-$.  In QCD, there are also the parity-conjugate tree level splitting amplitudes of the form $- \mapsto --$ and $+ \mapsto -+$.  These do not appear in our self-dual gauge theory, which only has a $++-$ vertex.   

\section{One loop corrections}
One loop QCD splitting amplitudes have been analyzed in \cite{Bern:1994zx, Kosower:1999rx, Bern:2005hs}. There are two new processes at one loop, namely the $- \mapsto ++$ amplitude and its parity conjugate. In the normalization of \cite{Bern:2005hs} this splitting amplitude is 
\begin{equation} 
	\op{Split}_+^{[1]} (a^+, b^+ ) = -\frac{N_c}{96 \pi^2}  \frac{[ab]}{\ip{ab}^2 } \label{eqn:splittingamplitude} 
\end{equation}

This is the only one-loop amplitude from the analysis of \cite{Bern:2005hs}  that contributes to self-dual gauge theory.  Indeed, all one-loop diagrams in self-dual gauge theory, when all particles are viewed as incoming, have positive helicity external lines. Therefore they can only contribute to a $- \mapsto ++$ splitting amplitude.  The remaining one-loop splitting amplitudes of QCD are either multiplicative corrections to the tree-level splitting amplitudes, or the parity-conjugate $- \mapsto --$ splitting amplitude.  Neither of these can appear from a one-loop diagram in self-dual gauge theory.

Perhaps surprisingly, our analysis will show that there are two other terms in the splitting amplitude of self-dual gauge theory, of the form $-+ \mapsto ++$ and $-- \mapsto -+$.   These splitting functions seem not to have appeared in other work on the topic; perhaps they are not visible in standard QCD amplitudes, but only in form factors.     

Let us implement the splitting amplitude \eqref{eqn:splittingamplitude} in the chiral algebra.  Looking at the definition of $[ab]$ and $\ip{ab}$ in the chiral algebra, we see that the only natural way is to add on a term in the OPE of the form
\begin{equation} 
	J_a[\til{\lambda}_1] (z_1) J_b[\til{\lambda}_2]  (z_2) \sim -\frac{N_c}{96 \pi^2} \frac{[12]}{\ip{12}^2} f_{ab}^c  \til{J}_c [\til{\lambda}_1 + \til{\lambda}_2] \left( \half z\right)    
\end{equation}
On the right hand side, the operator is evaluated at $\half z:= \frac{z_1 + z_2}{2}$; this is forced by symmetry.  We can rewrite this OPE as
\begin{equation}
	\begin{split} 	
		J_a[\til{\lambda}_1] (z_1) J_b[\til{\lambda}_2]  (z_2)  \sim & -\frac{N_c}{96 \pi^2} \frac{[12]}{\ip{12}^2} f_{ab}^c \til{J}_c [\til{\lambda}_1 + \til{\lambda}_2] (z_1) \\ & + \frac{N_c}{192 \pi^2} \frac{[12]}{\ip{12}}  f_{ab}^c \frac{1}{2 \pi \i} \partial_z \til{J}_c [\til{\lambda}_1 + \til{\lambda}_2] (z_1). 
\end{split}
\end{equation}
Finally, specializing the generating function $J[\til{\lambda}]$ to the terms linear in $\til{\lambda}$, we find the OPE
\begin{equation} 
	\begin{split} 
		J_a[1,0] (z_1) J_b[0,1]  (z_2)  \sim & \frac{N_c}{48 \pi^2} \frac{1}{ \ip{12}^2  } f_{ab}^c \til{J}_c [0,0] (z_1) \\ & - \frac{N_c}{96 \pi^2} \frac{1}{\ip{12}}  f_{ab}^c \frac{1}{2 \pi \i} \partial_z \til{J}_c [0,0] (z_1). 	 
	\end{split}
\end{equation}

\section{Failure of associativity of the corrected OPE} 
Now we can ask if this new OPE satisfies the associativity relation of a chiral algebra. The associativity of the OPE is encoded in equalities between contour integrals of OPEs involving three operators, coming from the change of contour.  For instance, one of the associativity equations is the identity
\begin{align}\nonumber
	&K^{ab} \oint_{\abs{z} =1, \abs{w} = 2}  J_a[1,0](0) J_b[0,1 ](z)  J_c[0,0 ](w) w \d w \d z \\
	&=  K^{ab}	\oint_{\abs{z}=2, \abs{w} = 1}  \left( J_a[1,0](0) J_c[0,0](w) w \d w \right) J_b[0,1 ](z)   \d z \\ \nonumber
	&+ K^{ab} \oint_{\abs{z} = 2, \abs{w} = 1} J_a[1,0](0) \\ \nonumber
	& \ \ \ \ \ \ \ \ \  \ \ \ \ \ \ \ \ \ \left( J_b[0,1 ](z)  J_c[0,0 ](w+z) (w+z)  \d w \right)  \d z.
	\label{eqn:associativity}	
\end{align}
where $K^{ab}$ is the Killing form.
We find that this identity fails to hold with our quantum corrected OPE \footnote{In this computation, as we will show shortly, there are no quantum corrections to OPEs involving $J[0,0]$ and $\tilde{J}[0,0]$; indeed, no such corrections are possible.}. 

We first note that the left hand side vanishes. If we take the $z$ contour integral first, the terms with a first order pole are anti-symmetric in the $a$ and $b$ indices.  Similarly,  the first term on the right hand side vanishes because there is no second order pole in $w$.  

For the second term  on the right hand, we can perform the $w$ contour integral first. This yields 
\begin{equation} 
	- \oint_{z} J_a[1,0](0) f_{bc}^d J_d[0,1] (z) z \d z  
\end{equation}
This is non-zero, because the one-loop correction to the OPE introduces a second-order pole.  The result of this contour integral is  
\begin{equation} 
	-\frac{1}{ 2 \pi \i } \frac{N_c}{48 \pi^2} \til{J}_e[0,0]K^{ab}  f_{ad}^e f_{bc}^d.    
\end{equation}
Since $K^{ab}  f_{ad}^e f_{bc}^d$ is the action of the quadratic Casimir in the adjoint representation, it is proportional to $\delta^{e}_c$.       We conclude that the one-loop corrected OPE is not associative.  

Alternatively, if we build a chiral algebra in which associativity is forced to hold, we find that all states of negative helicity become zero.

\section{Twistor space anomalies and chiral algebra associativity}
This failure of associativity is connected to the twistor space anomaly we have already mentioned.   As explained in \cite{Costello:2022wso}, from any local, anomaly-free theory on twistor space we can build a chiral algebra living on the twistor $\CP^1$.   

If we do this for the twistor uplift of self-dual gauge theory, then, at tree level, this matches the chiral algebra describing the tree-level collinear singularities in self-dual gauge theory. However, the twistor uplift is anomalous at loop level.  To cancel this anomaly, we need to introduce a new field on twistor space, corresponding to the axion in equation \eqref{eqn:axion}.  

We know on general grounds \cite{Costello:2022wso} that the theory including the axion corresponds to a consistent chiral algebra. Here, we will determine that this chiral algebra contains a one-loop correction to the classical OPE which matches the one-loop splitting amplitude.   

For associativity to hold, the axion field is essential. This tells us that the failure of associativity of the quantum-corrected OPE is a reflection of the twistor space anomaly, and is solved by the same Green-Schwarz mechanism.

To perform the calculation, we need to describe the extra elements in the chiral algebra coming from the axion field, and their OPEs.  Let us now do this.

\section{Chiral algebra including the axion}
The chiral algebra has four towers of states, each living in an infinite sum of finite-dimensional representations of $SU(2)$.  They are enumerated in Table \ref{table:chiralalgebra}.  In the chiral algebra presentation, we write the Lorentz group as $SU(2) \times SL_2(\R)$, where $SL_2(\R)$ rotates the chiral algebra plane.   Each state with label $m,n$ transforms in a representation of $SU(2)$ of highest weight $\half(m+n)$ and is a weight vector of weight $\half(m-n)$.   
\begingroup
\begin{table}
\begin{tabular}{c |  c |  c |  c   }
	Generator  & Spin &  Field & Dimension \\
	$J[m,n]$, $m,n \ge 0$ & $1-(m+n)/2$ &  $A$ & $-m-n$  \\
	$\til{J}[m,n]$, $m,n \ge 0$ & $-1-(m+n)/2$ &  $B$ & $-m-n-2$ \\
	$E[m,n]$, $m+n > 0$ &  $-(m+n)/2$ & $\rho$ & $-m-n$ \\
	$F[m,n]$, $m,n \ge 0$ & $ -(m+n)/2  $   & $\rho$ & $-m-n-2$  
\end{tabular}
	\caption{The generators of our 2d chiral algebra and their quantum numbers.  Dimension refers to the charge under scaling of $\R^4$.  \label{table:chiralalgebra}}
\end{table}
\endgroup
The OPEs involving the $E,F$ towers are 
\begin{multline}  
		 J^a[r,s](0) E[t,u](z) \\
		 \sim  \frac{1}{2 \pi \i z} \frac{(ts - ur)}{t + u}\what{\lambda}_\g \til{J}^a [t+r - 1, s + u -1](0)  
\end{multline}
\begin{multline} 
	 		 J^a[r,s](0) F[t,u](z)\\
			 \sim  -  \what{\lambda}_\g  \frac{1}{2 \pi \i z} \partial_z \til{J}^a[r+t, s + u](0) \\
		 -   \what{\lambda}_\g \frac{1}{2 \pi \i z^2} (1 + \frac{r + s}{t+u+2}) \til{J}^a[r+t, s+u](0) \\
\end{multline}
\begin{multline} 
	 		 J^a[r,s](0) J^b[t,u](z) \\ \sim \what{\lambda}_\g  \frac{1}{2\pi \i z} K^{ab} (ru-st) F[r+t-1,s+u-1] (0)\\
		 -  \what{\lambda}_\g \frac{1}{2\pi \i z} K^{ab} (t+u)  \partial_z E[r+t,s+u](0) \\
		 - \what{\lambda}_\g  \frac{1}{2\pi \i z^2} K^{ab} (r+s+t+u) E[r+t,s+u](0).  \label{eqn:EFope} 
\end{multline}
Let us explain the constant $\what{\lambda}_{\g}$. First, we define $\lambda_{\g}$ so that, for $X \in \g$, $\op{Tr}(X^4) = \lambda_{\g}^2 \op{tr}(X^2)^2$, where $\op{Tr}$ means the trace in the adjoint and $\op{tr}$ means the minimal trace (i.e.\ the fundamental for $SU(N)$).   This tensor identity is necessary for the Green-Schwarz mechanism to hold.   According to \cite{Okubo:1978qe}, we have 
\begin{equation} 
	\lambda_{\g}^2 = \frac{10 (\sh^\vee)^2 }{\dim \g + 2} 
\end{equation}
where $\sh^\vee$ is the dual Coxeter number, equal to $N_c$ for $SU(N_c)$.  

To account for the normalization of the interaction between the gauge field on twistor space and the field corresponding to the axion, we let
\begin{equation} 
	\what{\lambda}_{\g} = \frac{ \lambda_{\g} } { (2 \pi \i)^{3/2} \sqrt{12} }.
\end{equation}

\section{One loop corrections}

Let us now consider the possible one-loop corrections to the chiral algebra. We will then normalize them using associativity.   (In principle, one can compute these using the method of \cite{Costello:2022wso}, where the OPE coefficients are computed by an analysis of Feynman diagrams on twistor space. This was done in a similar situation for $5d$ gauge theories in \cite{Costello:2017fbo}, and a related analysis appears in the forthcoming work of \cite{Yehao}.   However, we find that fixing the OPE coefficients using associativity is significantly easier). 

Our chiral algebra has a symmetry called dimension in the Table \ref{table:chiralalgebra}.  This comes from scaling on $\R^4$. It is a symmetry that persists to the quantum level; therefore any OPE must respect this symmetry. 

One-loop corrections to the OPE cannot involve the axion, as axion exchanges are already a one-loop effect.  One-loop corrections are determined by the OPEs involving $J[0,0]$, $\til{J}[0,0]$, $J[1,0]$, $J[0,1]$.  This is because these generate the algebra.   Further, one-loop corrections must increase the number of $\til{J}$'s in an expression by one.  

The most general correction to the OPE between  $J[1,0]$, $J[0,1]$ currents is  
\begin{multline}
		J_a[1,0](0) J_b[0,1](z)\\
		= -\frac{1}{2 \pi \i z}   C K^{fe} (f_{ae}^c f_{bf}^d  + f_{ae}^d f_{bf}^c)   :J_c[0,0] \til{J}_d[0,0] :(0) \\
			+ \frac{1}{2 \pi \i z} \half  D f_{ab}^c    \partial_z \til{J}_c (0)
			+ \frac{1}{2 \pi \i z^{2}} D f_{ab}^c \til{J}_c(0).
\end{multline}
where $C,D$ are constants to be determined. The terms whose coefficient is $D$ correspond to the known one-loop splitting amplitudes of self-dual Yang-Mills.  

OPEs involving $J_a[0,0]$ cannot get deformed.  For instance, consider the OPE between $J_a[0,0]$ and $J_b[0,0]$. This must result in a state in the chiral algebra of dimension $0$ with exactly one copy of $\til{J}$. A glance at Table \ref{table:chiralalgebra} tells us is not possible.  A similar remark tells us that the OPE of $J_a[0,0]$ with any of the other states we are considering does not change. 

In addition to this OPE, associativity of the chiral algebra also forces a correction to the $J-\til{J}$ OPE, which is
\begin{equation}
	\begin{split} 
	J_a [1,0] (0)  \til{J}_b [0,1] (z)  \simeq C \frac{1}{z}  K^{fe} f_{ae}^c f_{bf}^d :\til{J}_c[0,0] \til{J}_d[0,0]:    \\
J_a [0,1] (0)  \til{J}_b [1,0] (z)  \simeq -C \frac{1}{z}  K^{fe} f_{ae}^c f_{bf}^d :\til{J}_c[0,0] \til{J}_d[0,0] :   
	\end{split}	
\end{equation}
for the same value of $C$ as above. 

\section{Normalizing the quantum corrections using associativity} 

The constants $C$ and $D$ in the quantum-corrected OPE will be chosen so that the following identity holds:
\begin{multline}
			\oint_{\abs{z} =1, \abs{w} = 2}  J_a[1,0](0) J_b[0,1 ](z)  J_c[0,0 ](w) w \d w \d z \\
			=  	\oint_{\abs{z}=2, \abs{w} = 1}  \left( J_a[1,0](0) J_c[0,0](w) w \d w \right) J_b[0,1 ](z)   \d z \\
	+   \oint_{\abs{z} = 2, \abs{z-w} = 1} J_a[1,0](0) \left( J_b[0,1 ](z)  J_c[0,0 ](w) w  \d w \right)  \d z. 
\label{eqn:associativity}	
\end{multline}
This identity is, of course, a consequence of the associativity of the OPE.    

\begin{proposition}
	The OPE associativity identity \eqref{eqn:associativity} holds if and only if we have the trace identity
	\begin{equation} 
		\op{Tr}(X^4) = \lambda_\g^2 \op{tr}(X^2)^2 
	\end{equation}
	and we take the constants $C,D$ to be 
\begin{equation} 
	\begin{split} 
		C & =  \frac{3}{2 (2\pi \i)^3 12  }	\\
		D &= -\frac{\sh^\vee}{ (2 \pi \i)^3 12  } 
	\end{split}	
	 \label{prop:OPEconstraint}
\end{equation}
\end{proposition}
The proof of this is provided in the supplementary materials.  It is an entirely routine argument: we simply compute both sides of equation \eqref{eqn:associativity} and compare them.  

\section{Matching coefficients}
To compare to the known results about one-loop splitting amplitudes, we should recall that $\ip{ij}$ comes with a factor of $2 \pi \i$ and $[ij]$ with a factor of $-2$. This brings the coefficient of the one-loop $++ \mapsto -$ term in the OPE to 
\begin{equation} 
	\frac{[ij]}{\ip{ij}^2}		\frac{ \sh^\vee}{(2 \pi \i)^2 24}   = -\frac{ \sh^\vee}{96 \pi^2}  \frac{[ij]}{\ip{ij}^2}. 
\end{equation}
For the groups $SU(N_c)$, $\sh^\vee = N_c$.  This matches the coefficient in \cite{Bern:2005hs}.  

We should emphasize that the chiral algebra, as an abstract chiral algebra, does not know about this constant. Multiplying the generators $\til{J}, E, F$ of the chiral algebra by a constant changes the coefficient of the one-loop OPE. In our analysis, however, we have taken care that our chiral algebra generators match states in the gauge theory exactly, without a prefactor. When we do this, we do find the correct coefficient, providing a chiral-algebraic derivation of the one-loop splitting amplitude.

\section{Amplitudes}
Associativity of the chiral algebra in the presence of the axion gives us a remarkably simple formula for the one-loop form factors of self-dual Yang-Mills in the presence of the operator $\half \op{tr}(B^2)$.  This form factor is the same as the one loop amplitude for QCD with an axion, with one particle of negative helicity.  We find that the form factor is 
\begin{equation}
	\begin{split} 	
		&\ip{\half \op{tr}(B^2) \middle|  1^- 2^+ \dots n^+ } \\=  	
		&\frac{1 }{ 192 \pi^2}  \sum_{2 \le i < j \le n} \frac{   [i j] \ip{1 i }^2  \ip{1 j }^2     }{\ip{ i j}  \ip{1 2} \ip{2 3} \dots \ip{n 1} } \op{Tr}_{\g}   \left(\t_{1} \dots \t_{n} \right) \\
		&+ \text{ permutations in } S_{n-1}	  \label{eqn:fullcorrelator1} 
\end{split}
\end{equation}
where we sum over permutations of the labels $2,\dots n$, and the trace is in the adjoint representation.  We note that this expression has \emph{both} the first order pole we found when the indices $i,j$ are not adjacent in the trace, as well as the second order pole when they are adjacent in the trace.   

This expression is valid also when the anomaly is cancelled by carefully chosen matter, instead of an axion, e.g.\ if the gauge group is $SU(2)$ and $N_f = 8$ (meaning $8$ fundamental and $8$ anti-fundamental).   In that case we must take the trace in the adjoint minus the matter representation. 

This formula is a great deal simpler than the formulae \cite{Mahlon:1993fe, Bern:1994zx} for the corresponding QCD amplitudes without an axion, bolstering our contention that the presence of the axion simplifies amplitudes greatly.  


\begin{acknowledgments}
We thank R. Bittleston, A. Sharma, and A. Strominger for helpful comments on a draft of this manuscript. K.C. is supported by the NSERC Discovery Grant program and by the Perimeter Institute for Theoretical Physics. Research at Perimeter Institute is supported by the Government of Canada through Industry Canada and by the Province of Ontario through the Ministry of Research and Innovation. N.P. acknowledges support from the University of Washington and the DOE award DE-SC0022347
\end{acknowledgments}

\bibliography{refs}
 \bibliographystyle{apsrev4-1}
\clearpage

\onecolumngrid

\pagebreak

\widetext
\clearpage
\begin{center}
\textbf{\large Supplemental Materials: \\ Associativity of the OPE with an axion }
\end{center}

\setcounter{equation}{0}
\setcounter{figure}{0}
\setcounter{table}{0}
\setcounter{page}{1}
\makeatletter
\renewcommand{\theequation}{S\arabic{equation}}
\renewcommand{\thefigure}{S\arabic{figure}}

In the supplemental material, we provide some technical details on a few points and give the derivation of Proposition \ref{prop:OPEconstraint}.  

\section{Normalization of states on twistor space } 
Give twistor space holomorphic coordinates $v_i, z$ related to coordinates on Euclidean $\R^4$ by  $v_1 = x_1 + \i x_2 + z (x_3 - \i x_4)$, $v_2 = x_3 + \i x_4 - z (x_1 - i x_2)$.   Our states $J[\til{\lambda}]$, $\til{J}[\til{\lambda}]$ are written in these coordinates as
\begin{equation} 
	\begin{split} 
		J[\til{\lambda}] (z_0) &= \delta_{z = z_0} e^{ \til{\lambda} \cdot v} \\
		\til{J}[\til{\lambda}] (z_0) &= \delta_{z = z_0} e^{ \til{\lambda} \cdot v}. 
	\end{split}
\end{equation}
When we write these expressions, we are trivializing the canonical bundle on the curve with coordinate $z$ using $\d z$. Further, since $\til{J}$ is associated to a field of negative helicity which lives in the canonical bundle of twistor space, we also have to trivialize this bundle which we do with the $3$-form $\d v_1 \d v_2 \d z$.

With this normalization, if we ask that the modes $\oint J_a[r,s] z^k \d z$ commute without a factor of $2 \pi \i$, then there must be a factor of $2 \pi \i$ present in the OPEs, so that
\begin{equation} 
	J_a [m,n](0) J_b [r,s](z) = -\frac{1}{2 \pi \i z} f_{ab}^c J_c[m+r, s+n] . 
\end{equation}
This factor can be absorbed by a rescaling of the currents $J[r,s]$, but we prefer not to here to match with a standard basis of states on momentum space. 

In \cite{Costello:2022wso, Costello:2021bah} there is a factor of $\what{\lambda}_{\g}$ in the interaction between the twistor uplift of the axion field and holomorphic BF theory on twistor space.  This leads to a corresponding factor in the OPEs which involve the $E,F$ towers, as in equations \eqref{eqn:EFope}.  This can be absorbed into a rescaling of $E,F$ and $\til{J}$, but here we prefer not to do that as it leads to a non-standard basis for negative helicity states.   

We normalize  $\ip{12} = 2 \pi \i (z_1 - z_2)$ so as to match standard conventions. We normalize the operator $\half \textrm{tr}(B^2)$ so that it gives the standard Parke-Taylor formula.    At a first pass, we can normalize $[12]$ so that $[12] = \tfrac{1}{2 \pi \i} \eps_{\dot \alpha \dot \beta} \til{\lambda}_{\dot \alpha} \til{\lambda}_{\dot \beta}$. 

With these normalizations we see that we will need to correct the normalization of $[ij]$ by a factor of $-2 \pi \i$.  To see this, we note that the first place that $[ij]$ appears is in the NMHV amplitude given by the CSW \cite{Cachazo:2004kj} formula.  There, two MHV vertices are glued together with a propagator which is that of a scalar field, namely $\norm{p}^{-2}$ in momentum space or $\tfrac{1}{4 \pi^2} \norm{x}^{-2}$ on real space.  However, on twistor space, the exchange of fields between two twistor lines acquires a different normalization.

According to the analysis of \cite{Cachazo:2004kj}, the factor of $\frac{1}{p^2}$ comes from the exchange of fields between operators supported on disjoint curves on twistor space. Place one operator $\int_{\CP^1} \mc{A}$  at the curve corresponding to $0$, and $\int_{\CP^1}\mc{B}$  at the curve corresponding to $x \in \R^4$. (Here $\mc{A}$, $\mc{B}$ are the fields of holomorphic BF theory). The field sourced by the operator placed at $0$ is $\frac{1}{2 \pi \i} \frac{1}{v_1} \delta_{v_2 = 0}$.  Integrating this over the curve at $x$ yields $\tfrac{1}{2 \pi \i} \norm{x}^{-2}$.  

This differs by a factor of $-2 \pi \i$ from the two-point function of a scalar field, $\tfrac{1}{4 \pi^2} \norm{x}^{-2}$.  This is fixed by multiplying $[12]$ by $-2 \pi \i$.  This leads us to a normalization where
\begin{equation} 
	\ip{12}[12] = p_1 \cdot p_2.  
\end{equation}
We further multiply $[12]$ by a factor of $2$ to get the final expression
\begin{equation} 
	[12] = -2 \eps_{\dot \alpha \dot \beta} \til{\lambda}_{\dot \alpha} \til{\lambda}_{\dot \beta} 
\end{equation}

\section{Proof of Proposition (\ref{prop:OPEconstraint}).} 
\subsection{The right-hand side of \eqref{eqn:associativity}. } First, let us compute the right-hand side of equation \eqref{eqn:associativity}.  There are two terms.  The first  is 
\begin{equation}
	\oint_{\abs{z}=2, \abs{w} = 1} \left(  J_a[1,0](0) J_c[0,0 ](w) w \d w \right)  J_b[0,1 ](z)   \d z .\label{eqn:RHS1}	
\end{equation}
Let us compute the $w$ integral first.
The only second order pole in the OPE of $J_a[1,0]$ with $J_c[0,0]$ is 
\begin{equation} 
	J_a[1,0] (0) J_c[0,0](w) \sim - \what{\lambda}_\g \frac{1}{2 \pi \i w^2} K_{ac} E[1,0](0) + O(1/w). 
\end{equation}
Then equation \eqref{eqn:RHS1} becomes
\begin{equation} 
	- \what{\lambda}_{\g} \oint_z K_{ac} E[1,0](0) J_b[0,1](z) \d z . 
\end{equation}
Next, we need to locate the first-order poles in the OPE between $E[1,0]$ and $J_b[0,1]$. We have
\begin{equation} 
	J_b[r,s](z) E[t,u](0) \sim  -   \what{\lambda}_\g \frac{1}{ 2 \pi \i z} \frac{(ts - ur)}{t + u} \til{J}_b [t+r - 1, s + u -1](0)   
\end{equation}
Therefore, \eqref{eqn:RHS1} evaluates to 
\begin{equation} 
	-\what{\lambda}_{\g} \oint_z K_{ac} E[1,0](0) J_b[0,1](z) \d z =\what{\lambda}_{\g}^2  K_{ac} \til{J}_b[0,0](0). 
\end{equation}

The second term on the right hand side of equation \eqref{eqn:associativity} is
\begin{equation} 
	\oint_{\abs{z} = 2, \abs{z-w} = 1} J_a[1,0](0) \left( J_b[0,1 ](z)  J_c[0,0 ](w) w  \d w \right) \d z.  
\end{equation}
Changing coordinates, this becomes 
\begin{equation} 
	\oint_{\abs{z} = 2, \abs{w} = 1} J_a[1,0](0) J_b[0,1 ](z)  J_c[0,0 ](w+z) (w+z)  \d w \d z.  
\end{equation}
Performing the $w$ contour integral, we pick up terms in the OPE between $J_b[0,1]$ and $J_c[0,0]$ which have either a first or a second order pole.   These are:
\begin{equation} 
	(w+z) J_b[0,1] (z) J_c[0,0](z+w) \simeq (w+z) \left( - \frac{1}{ 2 \pi \i w} f_{bc}^d  J_d[0,1](z) - \what{\lambda}_{\g} \frac{1}{ 2 \pi \i w^2} K_{bc}  E[0,1](z) \right)  
\end{equation}
Performing the contour integral over $w$ gives  
\begin{equation} 
-	z f_{bc}^d J_d[0,1](z) - \what{\lambda}_{\g}  K_{bc} E[0,1](z). 
\end{equation}
We need to perform the contour intgral
\begin{equation} 
	\oint_z J_a[1,0](0) \left( - z f_{bc}^d J_d[0,1](z)  -\what{\lambda}_{\g} K_{bc} E[0,1](z) \right) \d z \label{eqn:secondterm} 
\end{equation}
The OPEs which contribute to this contour integral are 
\begin{equation}
	\begin{split} 
		-J_a[1,0](0) J_d[0,1](z) \sim &   2 \what{\lambda}_{\g} \frac{1}{ 2 \pi \i z^2} K_{ad} E[1,1](0) \\
		& - f_{ad}^g D \frac{1}{ 2 \pi \i z^2} \til{J}_g[0,0] (0)\\
	-	\what{\lambda}_{\g} J_a[1,0](0) E[0,1](z) &\sim   \what{\lambda}_{\g}^2 \frac{1}{ 2 \pi \i z}  \til{J}_a [0,0](0)  
	\end{split}
\end{equation}

Therefore, the contour integral in equation \eqref{eqn:secondterm} yields 
\begin{equation} 
	\what{\lambda}_{\g}^2	\til{J}_a [0,0](0) K_{bc} + 2 \what{\lambda}_{\g}   K_{ad} f_{bc}^d E[1,1] (0) - f_{ad}^g f_{bc}^d  D \til{J}_g[0,0] (0) . 
\end{equation}
From this we see that the two terms in the right hand side of equation \eqref{eqn:associativity} sum to
\begin{equation} 
	\what{\lambda}_{\g}^2 K_{ac}\til{J}_b[0,0] + \what{\lambda}_{\g}^2  K_{bc} \til{J}_a [0,0]  + 2 \what{\lambda}_{\g}  K_{ad} f_{bc}^d E[1,1] 	- f_{ad}^g f_{bc}^d  D \til{J}_g[0,0] (0) . 
\end{equation}

\subsection{The left-hand side of \eqref{eqn:associativity}. }

The left-hand side of \eqref{eqn:associativity} is 
\begin{equation} 
	\oint_{\abs{z} =1, \abs{w} = 2}  \left( J_a[1,0](0) J_b[0,1 ](z) \d z \right) J_c[0,0 ](w) w \d w .\label{eqn:LHS} 
\end{equation}
We will perform the $z$ integral first.  To do this, we will write down all the terms in the OPE between $J_a[1,0](0)$ and $J_b[0,1](z)$ which have a first order pole.  These are:
\begin{equation}
	\begin{split} 
		-\frac{1}{ 2 \pi \i z}	f_{ab}^d J_d[1,1]  &+ 
		 \frac{1}{ 2 \pi \i z} K_{ab} \what{\lambda}_{\g} F[0,0] (0)
		- \frac{1}{ 2 \pi \i z} K_{ab} \what{\lambda}_{\g} \partial E[1,1](0)  \\
		&+ \frac{1}{ 2 \pi \i z} \half D f_{ab}^d \partial \til{J}[0,0](0)\\ 
		&-	C \frac{1}{ 2 \pi \i z}   K^{ij} f_{ai}^d   f_{bj}^e (: \til{J}_d[0,0] J_e[0,0]: + :\til{J}_e[0,0] J_d[0,0]:  )
	\end{split}
\end{equation}
where the last two terms are the quantum correction, and $C$, $D$ are constants we will determine.

After performing the $z$ integral, the left hand side of \eqref{eqn:associativity} becomes
\begin{equation} 
	 \begin{split} 
		 \oint_{w}-	f_{ab}^d J_d[1,1](0) J_c[0,0](w) w \d w  &+ 
		 \what{\lambda}_{\g}  K_{ab}  F[0,0] (0) J_c[0,0](w) w \d w\\
		 &-   \what{\lambda}_{\g}   K_{ab}  \partial E[1,1](0) J_c[0,0](w) w \d w  \\
&+ \half D f_{ab}^d \partial \til{J}_d[0,0] (0) J_c[0,0](w) w \d w  \\ 
		 &-	C    K^{ij} f_{ai}^d   f_{bj}^e (: \til{J}_d[0,0] J_e[0,0]: \\
		 & + :\til{J}_e[0,0] J_d[0,0]:  ) (0) J_c[0,0](w) w \d w 
	\end{split} \label{eqn:intermediate_OPE}
\end{equation}
The first three terms are straightforward to evaluate, as they do not involve composite operators.   In each term, we are only interested in OPEs with poles of order $2$, because we are integrating against $w \d w$.   These are:
\begin{equation} 
	\begin{split}
		J_d[1,1](0) J_c[0,0](w) &\sim -  \what{\lambda}_{\g}   \frac{1}{ 2 \pi \i  w^2} 2 E[1,1] (0) K_{dc} \\ 
  F[0,0](0) J_c[0,0](w) &\sim  -   \what{\lambda}_{\g}   \frac{1}{ 2 \pi \i w^2} \til{J}_c[0, 0](0)
	\end{split}
\end{equation}
The third term in equation \eqref{eqn:intermediate_OPE} does not have any non-singular OPEs. 

The fourth term in \eqref{eqn:intermediate_OPE} is 
\begin{equation} 
	 \oint_w  \half D f_{ab}^d \partial \til{J}_d[0,0] (0) J_c[0,0](w) w \d w   
\end{equation}
We have the OPE
\begin{equation} 
	\partial \til{J}_d[0,0]( 0 ) J_c[0,0] (w) \sim -f_{dc}^e \til {J}_e[0,0] (0) \frac{1}{ 2 \pi \i w^2} 
\end{equation}
so that this contour integral is
\begin{equation} 
-	\half D f_{ab}^d f_{dc}^e \til{J}_e[0,0] (0). 
\end{equation}
Putting this together, we find the first four terms of the contour integral of \eqref{eqn:intermediate_OPE} we are computing are
\begin{equation} 
	2 \what{\lambda}_{\g} E[1,1]  f_{ab}^d K_{cd} -  \what{\lambda}_{\g}^2 \til{J}_c[0,0] K_{ab} - \half D f_{ab}^d f_{dc}^e \til{J}_e[0,0] (0).  
\end{equation}

The next term to compute is
\begin{equation} 
	-C   \oint_w  K^{ij} f_{ai}^d   f_{bj}^e (: \til{J}_d[0,0] J_e[0,0]: + :\til{J}_e[0,0] J_d[0,0]:  )(0)  J_c[0,0](w) w \d w. \label{eqn:NO_contour} 
\end{equation}
This can be written in terms of double contour integrals. For example, the first term is
\begin{equation} 
	-C \oint_{\abs{z} = 1, \abs{w} = 2}  \til{J}_d[0,0](0) J_e[0,0](z) J_c[0,0](w) \frac{\d z}{ 2 \pi \i z} w \d w. 
\end{equation}
Changing the contour changes this to
\begin{equation} 
	-C\oint_{\abs{z} = 2, \abs{w} = 1}  \til{J}_d[0,0](0) J_e[0,0](z) J_c[0,0](w+z) \frac{\d z}{ 2 \pi \i z} (w+z) \d w .\label{eqn:contour} 
\end{equation}
The other contour, where $J_c[0,0](w)$ moves around $\til{J}_d[0,0](0)$, does not contribute.  This is because there is only a first order pole in the $J[0,0]$--$\til{J}[0,0]$ OPE. 

The only OPE that contributes to the $w$ integral in equation \eqref{eqn:contour} is the Kac-Moody OPE
\begin{equation} 
	J_e[0,0](z) J_c[0,0](w+z) \sim -\frac{1}{ 2 \pi \i w} f_{ec}^f J_f[0,0]. 
\end{equation}
Therefore,  the $w$ integral in  \eqref{eqn:contour} evaluates to 
\begin{equation} 
	C \oint_{z}  \til{J}_d[0,0](0) f_{ec}^f J_f[0,0](z) \d z 
\end{equation}
which can be readily evaluated to give
\begin{equation} 
	 -C  \til{J}_g[0,0]  f_{ec}^f  f_{df}^g.   
\end{equation}
From this we see that the contour integral \eqref{eqn:NO_contour}   becomes
\begin{equation}
-	C \til{J}_g[0,0]    (  K^{ij} f_{ai}^d   f_{bj}^e   f_{ec}^f f_{df}^g   +    K^{ij}  f_{bj}^d  f_{ai}^e   f_{ec}^f f_{df}^g  )  
\end{equation}
Let us change this expression by lowering all indices using the Killing form, and contracting repeated indices using the Killing form. It becomes
\begin{equation}
	-C \til{J}_g[0,0]    (   f_{aid}   f_{bie}   f_{ecf} f_{dfg}   +      f_{bid}  f_{aie}   f_{ecf} f_{dfg}  )  
\end{equation}
Rearranging and permuting the indices this becomes
\begin{equation}
  	C \til{J}_g[0,0]    (   f_{adi}   f_{bie}   f_{cef} f_{gfd}   +      f_{bdi}   f_{aie}   f_{cef} f_{gfd}    )  
\end{equation}

Writing this in terms of a basis $\t^a$ of the Lie algebra, we find that this is 
\begin{equation}
	C \til{J}_g[0,0] \left(\op{Tr}( \t_a \t_b \t_c \t_g ) + \op{Tr}(\t_b \t_a \t_c \t_g) \right) \label{eqn:trace} 
\end{equation}
where the trace is taken in the adjoint representation. 

Putting this together, we find that the left hand  of equation \eqref{eqn:associativity} is
\begin{equation} 
	2\what{\lambda}_{\g} E[1,1]  f_{abc} -\what{\lambda}_{\g}^2 \til{J}_c[0,0] K_{ab}   	+C \til{J}_g[0,0] \left(\op{Tr}( \t_a \t_b \t_c \t_g ) + \op{Tr}(\t_b \t_a \t_c \t_g) \right) -  \half D f_{ab}^d f_{dc}^g \til{J}_g[0,0] (0).   
\end{equation}
It is natural to write the last term in terms of commutators as well, so we get 
\begin{equation} 
	2\what{\lambda}_{\g} E[1,1]  f_{abc} -\what{\lambda}_{\g}^2 \til{J}_c[0,0] K_{ab}   	+C \til{J}_g[0,0] \left(\op{Tr}( \t_a \t_b \t_c \t_g ) + \op{Tr}(\t_b \t_a \t_c \t_g) \right) +  \half D \op{tr}\left( [\t_c, [\t_a,\t_b]] \t_{\g}\right)   \til{J}_g[0,0] (0).   
\end{equation}
Finally, we can move from the trace in the fundamental to the trace in the adjoint at the price of a copy of twice the dual Coxeter number $2\sh^\vee$, giving us 
\begin{equation} 
	2\what{\lambda}_{\g} E[1,1]  f_{abc} -\what{\lambda}_{\g}^2 \til{J}_c[0,0] K_{ab}   	+C \til{J}_g[0,0]  \left(\op{Tr}( \t_a \t_b \t_c \t_g ) +  \op{Tr}(\t_b \t_a \t_c \t_g) \right) + \frac{1}{4 \sh^\vee} D \op{Tr}\left( [\t_c, [\t_a,\t_b]] \t_{\g}\right)   \til{J}_g[0,0] (0).   
\end{equation}

\subsubsection{Comparing the left and right hand sides}
The right hand side is
\begin{equation} 
	K_{ac} \what{\lambda}_{\g}^2 \til{J}_b[0,0] + K_{bc} \what{\lambda}_{\g}^2 \til{J}_a [0,0]  + 2  f_{bca} E[1,1] -  f_{ad}^g f_{bc}^d  D \til{J}_g[0,0] (0) . 	
\end{equation}
We can rewrite the last term as
\begin{equation} 
-	\op{tr}\left([\t_a,[\t_b,\t_c]] \t_{g}  \right)     D \til{J}_g[0,0] (0)  = -\frac{1}{2\sh^\vee} 	\op{Tr}\left([\t_a,[\t_b,\t_c]]  \t_{g} \right)     D \til{J}_g[0,0] (0) 
\end{equation}
where $\sh^\vee$ is the dual Coxeter number.  

Comparing the right and left hand sides, we see that the coefficients of the operator $E[1,1]$ match.  Removing this operator, we find that the associativity equation \eqref{eqn:associativity} holds if
\begin{multline} 
	 -  \what{\lambda}_{\g}^2 \til{J}_c[0,0] K_{ab}   	+C \til{J}_d[0,0] \left(\op{Tr}( \t_a \t_b \t_c \t_d ) + \op{Tr}(\t_b \t_a \t_c \t_d) \right)  +\half D \tfrac{1}{2\sh^\vee}  \op{Tr}\left( [\t_c, [\t_a,\t_b]] \t_{\d}\right)   \til{J}_d[0,0] (0)   \\
	=  \what{\lambda}_{\g}^2  K_{ac} \til{J}_b [0,0] +   \what{\lambda}_{\g}^2  K_{bc} \til{J}_a [0,0]  - D \tfrac{1}{2\sh^\vee} 	\op{Tr}\left([\t_a,[\t_b,\t_c]] \t_{d} \right)     \til{J}_d[0,0] (0)  	 
\end{multline}
This holds if the following Lie algebra identity holds: 
\begin{multline}
	C  \left(\op{Tr}( \t_a \t_b \t_c \t_d ) + \op{Tr}(\t_b \t_a \t_c \t_d) \right)  -\half D \tfrac{1}{2\sh^\vee}  \op{Tr}\left(  [\t_a,\t_b][\t_c, \t_{\d}]\right)   - D \tfrac{1}{2\sh^\vee}	\op{Tr}\left([\t_b,\t_c] [\t_a, \t_d] \right)     \\
	=  \what{\lambda}_{\g}^2 \left( K_{ac} K_{bd} + K_{ab} K_{cd} + K_{ad} K_{bc} \right).\label{eqn:Lie_algebra} 
\end{multline}
Here we have rewritten the terms involving commutators slightly, for later convenience.

Let us verify this using the trace identity we need to cancel the anomaly.  The first thing to note is that, since the adjoint representation is a real representation (i.e.\ equipped with an invariant symmetric pairing) the trace in the adjoint representation is anti-invariant under the dihedral group, not just the cyclic group.  This implies that the expression
\begin{equation} 
	\op{Tr}(\t_a \t_b \t_c \t_d) +  \op{Tr}(\t_a \t_c \t_d \t_b) +  \op{Tr}(\t_a \t_d \t_b \t_c)   
\end{equation}
is totally symmetric.  By our trace identity, we have
\begin{equation} 
	\op{Tr}(\t_a \t_b \t_c \t_d) +  \op{Tr}(\t_a \t_c \t_d \t_b) +  \op{Tr}(\t_a \t_d \t_b \t_c)   = \lambda_{\g}^2  \left( K_{ac} K_{bd} + K_{ab} K_{cd} + K_{ad} K_{bc} \right).  
\end{equation}
where $\what{\lambda}_{\g} = \frac{\lambda_{\g} } { (2 \pi \i)^{3/2} \sqrt{12} }$.

To verify the Lie algebra identity, we need to match the terms on the left hand side of equation \eqref{eqn:Lie_algebra} that are not totally symmetric.  We have
\begin{equation} 
	\begin{split} 
		3   \left(\op{Tr}( \t_a \t_b \t_c \t_d ) + \op{Tr}(\t_b \t_a \t_c \t_d) \right) & - 2 \left( \op{Tr}(\t_a \t_b \t_c \t_d) +  \op{Tr}(\t_a \t_c \t_d \t_b) +  \op{Tr}(\t_a \t_d \t_b \t_c)  \right)  \\
		& =\op{Tr}( \t_a \t_b \t_c \t_d) + \op{Tr}(\t_a \t_c  \t_d \t_b) - 2 \op{Tr}(\t_a \t_d \t_b \t_c) \\
		&= \op{Tr}([\t_d,\t_a]  \t_b \t_c) + \op{Tr}([\t_a,\t_c] \t_d \t_b ) .	
	\end{split}
\end{equation}
By dihedral symmetry we have
\begin{equation} 
	\op{Tr}([\t_d,\t_a]  \t_b \t_c) =  \tfrac{1}{2} \op{Tr}([\t_d,\t_a]  [\t_b, \t_c]) 
\end{equation}
so that
\begin{equation} 
	\begin{split} 
		 3   \left(\op{Tr}( \t_a \t_b \t_c \t_d ) + \op{Tr}(\t_b \t_a \t_c \t_d) \right) & - 2 \left( \op{Tr}(\t_a \t_b \t_c \t_d) +  \op{Tr}(\t_a \t_c \t_d \t_b) +  \op{Tr}(\t_a \t_d \t_b \t_c)  \right)  \\
		&= \tfrac{1}{2} \op{Tr}([\t_d,\t_a]  [\t_b, \t_c]) +  \half \op{Tr}([\t_a,\t_c] [\t_d ,\t_b]) \\
		&= -\tfrac{1}{2} \op{Tr}([\t_a,\t_d]  [\t_b, \t_c]) +  \half \op{Tr}([\t_b, [\t_a,\t_c]] \t_d ) \\
		&= -\tfrac{1}{2} \op{Tr}([\t_a,\t_d]  [\t_b, \t_c]) +  \half \op{Tr}( [[\t_b,\t_a],\t_c] \t_d )  +   \half \op{Tr}( [\t_a,[\t_b,\t_c]] \t_d )    \\
		&= - \op{Tr}([\t_a,\t_d]  [\t_b, \t_c]) +  \half \op{Tr}( [[\t_b,\t_a],\t_c] \t_d ) \\
		&= - \op{Tr}([\t_a,\t_d]  [\t_b, \t_c]) -  \half \op{Tr}( [\t_b,\t_a][\t_c, \t_d] ) .
	\end{split}
\end{equation}

Therefore we have
\begin{equation} 
	\begin{split} 
		\tfrac{3}{2}   \left( \op{Tr}( \t_a \t_b \t_c \t_d ) + \op{Tr}(\t_b \t_a \t_c \t_d) \right)& - \lambda_{\g}^2 ( K_{ac} K_{bd} + K_{ab} K_{cd} + K_{ad} K_{bc} )\\
		&	= - \frac{1}{2 } \op{Tr}([\t_a,\t_d]  [\t_b, \t_c]) -  \frac{1}{4 }  \op{Tr}( [\t_b,\t_a][\t_c, \t_d] ) .
	\end{split}
\end{equation}
Therefore, the Lie algebra identity \eqref{eqn:Lie_algebra} holds if we take
\begin{equation} 
	\begin{split} 
		C & =  \frac{3}{2 (2\pi \i)^3 12  }	\\
		D &= -\frac{\sh^\vee}{ (2 \pi \i)^3 12  } 
	\end{split}
\end{equation}

\section{Computing form factors}

Now we will carefully derive our equation \eqref{eqn:fullcorrelator1}  for form factors. Our starting point is that the tree level form factor is given by the Parke-Taylor formula 
\begin{equation} 
	\ip{\half \op{tr}(B^2) \mid \til{J}^{a_1}[\til{\lambda}_1] (z_1)  J^{a_2} [\til{\lambda}_2](z_2) \dots  \til{J}^{a_i}[\til{\lambda}_i](z_i) \dots  J^{a_n} [\til{\lambda}_n](z_n) } =   \sum_{\sigma \in S_{n-1}} \frac{ \ip{1i}^4}{\ip{\sigma_1 \sigma_2} \dots \ip{\sigma_n \sigma_1} } \op{tr}( \t_{a_{\sigma_1} } \dots \t_{a_{\sigma_n} } ).  
\end{equation}
Our target expression for the one loop form factor is 
\begin{equation} 
	 \frac{- \sh^\vee }{ 96 \pi^2}  \sum_{2 \le i < j \le n} \frac{   [i j] \ip{1 i }^2  \ip{1 j }^2     }{\ip{ i j}  \ip{1 2} \ip{2 3} \dots \ip{n 1} } \op{Tr}_{\g}   \left(\t_{1} \dots \t_{n} \right). 
\end{equation}
Let us rewrite this slightly.  We let $J[1] = J[1,0] \til{\lambda}^{\dot 1} + J[0,1] \til{\lambda}^{\dot 2}$, and similarly for $\til{J}[1]$, and we use the shorthand notation $J$, $\til{J}$ for $J[0,0]$ and $\til{J}[0,0]$. 

In this notation we aim to prove the two identities:
\begin{multline} 
	\ip{\half \op{tr}(B^2) \middle| \til{J}^{a_1}(z_1) J_{a_2}[1](z_2) J_{a_3}[1](z_3) J_{a_4}(z_4) \dots J_{a_n}(z_n) }  \\
	= -\frac{1}{n} \sum_{\sigma \in S_n}  \frac{   [23]\ip{12}^2  \ip{13}^2     }{\ip{23}  \ip{\sigma_1 \sigma_2} \ip{\sigma_2 \sigma_3} \dots \ip{\sigma_n \sigma_1} } \op{Tr}_{\g} \left(\t_{a_{\sigma_1}} \dots \t_{a_{\sigma_n}} \right) \label{eqn:correlator1}  
\end{multline}
\begin{equation} 
		\ip{\half \op{tr}(B^2) \middle| \til{J}^{a_1}[1] (z_1) J_{a_2}[1](z_2) J_{a_3}(z_3) J_{a_4}(z_4) \dots J_{a_n}(z_n) }= 0.\label{eqn:correlator2} 
\end{equation}
We will prove both of these by induction on the number of insertions. 

First, we will check that one-loop correlator factors with two insertions vanish.  The only possible non-zero one-loop correlator with two insertions is $\ip{\half \op{tr}(B^2) \mid \til{J}[1] (z_1) J[1](z_2)}$ (if we have a $J[2]$ or $\til{J}[2]$ insertion we get zero for symmetry reasons).  The OPE between $\til{J}[1] (z_1) J[1](z_2)$ is proportional to
\begin{equation} 
	\frac{[12]}{\ip{12}} : \til{J} \til{J} : (z_1). 
\end{equation}
The tree-level correlator with insertions $\til{J}(z_1) \til{J}(z_3)$ is $\ip{13}^2$. Therefore the one-point function of the normally ordered operator $: \til{J} \til{J}:$ vanishes, as it is the limit of this expression as $z_3 \to z_1$. 

Next, let us check the three point correlators.  It is easy to see that
\begin{equation} 
	\ip{\half \op{tr}(B^2) \middle| \til{J}[1](z_1) J[1](z_2) J(z_3) } = 0. 
\end{equation}
Indeed, the poles in this expression at $z_3 = z_2$, $z_3 = z_1$ are two-point correlators that we have already seen must vanish. In the same way, by induction, we find that \eqref{eqn:correlator2} holds.

The remaining three-point correlator is
\begin{equation} 
	\ip{\half \op{tr}(B^2) \middle| \til{J}_{a_1}(z_1) J_{a_2}[1](z_2) J_{a_3}[1](z_3) }. 
\end{equation}
As a function of $z_3$, this has poles at $z_1,z_2$ dictated by one-loop and tree-level OPEs.  The terms corresponding to tree-level OPEs vanish, as they involve a one-loop two-point function which vanishes.   There is a one-loop OPE when $z_3 = z_2$. 

The only way to contract the Lie algebra indices is by $f_{a_1 a_2 a_3}$, which is anti-symmetric under permutation of $2,3$.  Therefore we need only consider those OPEs at $z_2 = z_3$ which are anti-symmetric, i.e. involve an odd number of the symbols $[23]$, $\ip{23}$.  The only such pole is 
\begin{equation} 	
		J_{a_2}[1] (z_2) J_{a_3}[1](z_3) =		- \frac{[23] \sh^\vee }{ \ip{23}^2 96 \pi^2 }  f_{a_2 a_3}^c \til{J}_c(z_2).	
\end{equation}
Therefore we have
\begin{equation} 	
	\ip{\half \op{tr}(B^2) \middle| \til{J}_{a_1}(z_1) J_{a_2}[1](z_2) J_{a_3}[1](z_3) } \sim - \frac{[23] \ip{12}^2  \sh^\vee} { \ip{23}^2 96 \pi^2 } f_{a_1 a_2 a_3} 	
\end{equation}
where $\sim$ means we are considering the polar part at $z_2 = z_3$. The unique global expression with the correct poles and zeroes is
\begin{equation}
	\begin{split} 
		\ip{\half \op{tr}(B^2) \middle| \til{J}_{a_1}(z_1) J_{a_2}[1](z_2) J_{a_3}[1](z_3) } &= - \frac{[23] \ip{12} \ip{13}  \sh^\vee} { \ip{23}^2 96 \pi^2 } f_{a_1 a_2 a_3} \\
		&= \frac{1}{96 \pi^2} \frac{[23] \ip{12}^2 \ip{13}^2  \sh^\vee }{ \ip{23}  \ip{12} \ip{23} \ip{31} } f_{a_1 a_2 a_3}  
		\end{split}
\end{equation}

We can trade twice the dual Coxeter number for a trace in the adjoint representation, giving
\begin{equation} 
	\ip{\half \op{tr}(B^2) \middle| \til{J}_{a_1}(z_1) J_{a_2}[1](z_2) J_{a_3}[1](z_3) } =  \frac{1}{3} \frac{[23] \ip{12}^2 \ip{13}^2}   { 192 \pi^2  \ip{23} }  \sum_{\sigma \in S_3}\frac{1}{  \ip{\sigma_1 \sigma_2}\ip{\sigma_2 \sigma_3}  \ip{\sigma_3 \sigma_1}  }\op{Tr}_{\g} ( \t_{a_{\sigma_1}} \t_{a_{\sigma_2}} \t_{a_{\sigma_3}} ). 
\end{equation}
Then, proceeding inductively  using  the tree-level OPEs involving the last insertion, we find that 
\begin{multline} 	
	\ip{\half \op{tr}(B^2) \middle| \til{J}^{a_1}(z_1) J_{a_2}[1](z_2) J_{a_3}[1](z_3) J_{a_4}(z_4) \dots J_{a_n}(z_n) } \\
	= \frac{1}{192 \pi^2} \frac{1}{n}  \frac{   [23]\ip{12}^2  \ip{13}^2     }{\ip{23}}  \sum_{\sigma \in S_n} \frac{1}{\ip{\sigma_1 \sigma_2} \ip{\sigma_2 \sigma_3} \dots \ip{\sigma_n \sigma_1} } \op{Tr}_{\g } \left(\t_{a_{\sigma_1}} \dots \t_{a_{\sigma_n}} \right) \label{eqn:correlator1} 
\end{multline}

\end{document}